\documentclass[
reprint,
nofootinbib,
amsmath,amssymb,
aps,
onecolumn
]{revtex4-2}

\DeclareUnicodeCharacter{04E7}{\"o}

\usepackage{graphicx}
\usepackage{dcolumn}
\usepackage{bm}
\usepackage{color}
\usepackage{comment}
\usepackage{hyperref}
\usepackage{url}
\usepackage{amsmath}

\begin{document}

\title{\texorpdfstring{Generalized Hamiltonian formalism \\ for spatially nonlocal nonlinear differential equations}{Generalized Hamiltonian formalism for spatially nonlocal nonlinear differential equations}}

\author{Ali Pazarci}
 \affiliation{Department of Physics, Bogazici University,
34342 Bebek, Istanbul, Türkiye}

\author{Nadir Ghazanfari}
\affiliation{Department of Electrical and Electronics Engineering, Istinye University,
34485 Istanbul, Türkiye}

\author{Ilmar Gahramanov}
\affiliation{Department of Physics, Bogazici University,
34342 Bebek, Istanbul, Türkiye}%
\affiliation{Center for Mathematics and its Apllications, Khazar University,
Mehseti St. 41, AZ1096, Baku, Azerbaijan}

\date{\today}

\begin{abstract}
In this work, we develop a generalized Hamiltonian formalism for spatially nonlocal field theories whose Lagrangian densities depend explicitly on both the local field and its spatially reflected counterpart. Starting from a generalized variational principle, we derive generalized Euler-Lagrange equations and introduce a generalized functional derivative that consistently accounts for reflected-field contributions. The proposed formalism is applied to three spatially nonlocal nonlinear Schr\"odinger equations. For the Ablowitz-Musslimani equation, we construct, to the best of our knowledge, the first standard Lagrangian density formulated directly in terms of the complex fields and derive its complete Hamiltonian formulation. The same framework is subsequently applied to the two Lagrangian nonlocal nonlinear Schr\"odinger equations introduced by Velasco-Juan and Fujioka, yielding consistent Hamiltonian formulations that reproduce the corresponding generalized Euler-Lagrange equations. These results establish a unified Hamiltonian framework for a broad class of spatially nonlocal nonlinear field theories.  
\end{abstract}

\maketitle

\section{Introduction}

The nonlinear Schr\"odinger (NLS) equation is one of the most important integrable nonlinear evolution equations in mathematical physics, used to describe the behavior of physical systems ranging from the nonlinear optics~\cite{Powers2017, Agrawal2019} and classical fluid mechanics~\cite{hoefer2016, Kharif2025}
to quantum gases such as Bose-Einstein condensates~\cite{Pethick2008, Pitaevskii2016}. In addition to its wide application areas, the NLS equation also possesses well-established Lagrangian and Hamiltonian formulations~\cite{DERIGLAZOV2009, Pazarci2023}, making it a fundamental model in the study of integrable systems and constrained Hamiltonian dynamics~\cite{zakharov-shabat-1972-jetp,Zakharov:1974zf, belyutin1997, prinari2023}.

In 2013, Ablowitz and Musslimani~\cite{ablowitz2013integrable} introduced an integrable spatially nonlocal generalization of the NLS equation in which the nonlinear interaction couples the field at the point $x$ to its complex conjugate evaluated at the reflected point $-x$. This equation led to a rapidly growing area of research on nonlocal integrable systems, leading to the introduction of numerous spatially, temporally, and space-time nonlocal nonlinear equations, including nonlocal modified Korteweg-de Vries ~\cite{Ablowitz2016}, Hirota~\cite{Fring2019}, and Fordy-Kulish~\cite{Gurses2017} equations.

Although the integrability properties of nonlocal equations have been studied extensively~\cite{ablowitz2017, Ablowitz2018,ablowitz2021}, their variational and Hamiltonian structures remain much less understood. Fujioka and Espinosa~\cite{fujioka2019} showed that nonlocal dependence on spatially translated fields leads to a modified Euler-Lagrange equation. More recently, Velasco-Juan and Fujioka~\cite{velasco2022} introduced two nonlocal nonlinear Schrödinger equations, denoted as LN1 and LN2, which possess standard real Lagrangian densities. They also argued that a standard real Lagrangian does not exist for the Ablowitz-Musslimani system, although Ablowitz and Musslimani had previously demonstrated that the equation admits a Hamiltonian formulation~\cite{Ablowitz2018}.

In this work, we show that a real Lagrangian also exists for the Ablowitz-Musslimani (AM) equation and construct a consistent Hamiltonian formulation for all three nonlocal NLS systems. To accomplish this, we develop a generalized Hamiltonian formalism for spatially nonlocal field theories. Starting from a generalized variational principle, we derive generalized Euler-Lagrange equations that are appropriate for Lagrangian densities depending simultaneously on the fields evaluated at $x$ and their spatially reflected counterparts at $-x$. This naturally leads to a generalized functional derivative, which in turn allows the canonical Poisson bracket to be extended to spatially nonlocal systems while preserving the algebraic structure of the conventional Hamiltonian formalism. 

The paper is organized as follows. In Section~\ref{NL formalism}, we derive the generalized Euler-Lagrange equations and formulate the generalized Hamiltonian formalism for spatially nonlocal field theories. Sections~\ref{AM formalism}-\ref{LN2 formalism} apply the formalism to three spatially nonlocal NLS systems, namely the AM, LN1, and LN2 equations, respectively. For the AM equation, we construct, to the best of our knowledge, the first standard Lagrangian density written directly in terms of the complex fields and develop its complete Hamiltonian formulation using the generalized Dirac-Bergmann algorithm. The same framework is then applied to the LN1 and LN2 equations, demonstrating that the proposed formalism provides a unified Hamiltonian description of different classes of spatially nonlocal NLS equations. Finally, Section~\ref{Discussion} summarizes the main results and discusses possible extensions of the present framework.

\section{Hamiltonian Formalism for Nonlocal Field Theory}\label{NL formalism}

Nonlocal dependence in a Lagrangian density can require a modification of the conventional Euler-Lagrange equations. Fujioka and Espinosa ~\cite{fujioka2019} demonstrated that for a Lagrangian density depending on spatially translated fields, the Euler-Lagrange equation involves contributions evaluated at their shifted positions. In the present work, we instead develop a generalized formalism for the Euler-Lagrange equation of spatially nonlocal fields, which naturally leads to the definition of generalized functional derivatives, providing the foundation for the generalized Hamiltonian formalism.

We consider an arbitrary field $u(x,t)$ whose action functional is given by
\begin{equation}
S[u]= \iint dt \, dx\, \mathcal{L},
\label{eq:action}
\end{equation}
where the Lagrangian density is allowed to depend explicitly on both the field evaluated at the spatial point $x$ and its reflected point $-x$. In its most general form,
\begin{align}
\mathcal{L} = \mathcal{L}\Big(
& u(x,t), u_t(x,t), u_x(x,t), u_{2x}(x,t), \ldots, u_{nx}(x,t), \nonumber\\
& u(-x,t), u_t(-x,t), u_x(-x,t), u_{2x}(-x,t), \ldots, u_{nx}(-x,t) \Big).
\label{eq:generalLagrangian}
\end{align}
Here, $u_{kx}\equiv\partial^k u/\partial x^k$ denotes the $k$th spatial derivative. The presence of the reflected fields distinguishes the action Eq.~\eqref{eq:action} from its local counterpart such that the variation of the action receives contributions from the fields evaluated at $x$ and $-x$. The first variation of the action gives
\begin{align}
\delta S = \iint_{-\infty}^{\infty} dx\,dt \Bigg[
& \frac{\partial\mathcal{L}}{\partial u(x,t)} \,\delta u(x,t) + \frac{\partial\mathcal{L}}{\partial u_t(x,t)} \,\delta u_t(x,t) + \sum_{k=1}^{n} \frac{\partial\mathcal{L}} {\partial u_{kx}(x,t)} \,\delta u_{kx}(x,t) \nonumber\\
& + \frac{\partial\mathcal{L}}{\partial u(-x,t)} \,\delta u(-x,t) + \frac{\partial\mathcal{L}}{\partial u_t(-x,t)} \,\delta u_t(- x,t) + \sum_{k=1}^{n} \frac{\partial\mathcal{L}} {\partial u_{kx}(-x,t)} \,\delta u_{kx}(-x,t)
\Bigg].
\label{eq:firstvariation}
\end{align}
The variations satisfy
\begin{align}
\delta u_t(x,t) &= \frac{\partial}{\partial t}\delta u(x,t), \\
\delta u_{kx}(x,t) &= \frac{\partial^k}{\partial x^k}\delta u(x,t), \\
\delta u_t(-x,t) &= \frac{\partial}{\partial t}\delta u(-x,t), \\ 
\delta u_{kx}(-x,t) &= (-1)^k \frac{\partial^k}{\partial x^k} \delta u(-x,t),
\end{align}
where the factor $(-1)^k$ follows from repeated application of the chain rule.
Substituting the relations for the field variations into Eq.~\eqref{eq:firstvariation}, integrating by parts with respect to the temporal and spatial coordinates, and assuming that the boundary terms vanish, we obtain
\begin{align}
\delta S = \iint_{-\infty}^\infty  dx\,dt \, \mathcal{E}_{x}(u) \, \delta u(x,t) 
+ \iint_{-\infty}^\infty dx\,dt \, \mathcal{E}_{-x}(u) \, \delta u(-x,t),
\label{eq:variationSeparated}
\end{align}
where
\begin{align}
\mathcal{E}_{x}(u) = 
\frac{\partial\mathcal{L}}{\partial u(x,t)} - \frac{\partial}{\partial t} \left( \frac{\partial\mathcal{L}} {\partial u_t(x,t)} \right) 
+ \sum_{k=1}^{n} (-1)^k \frac{\partial^k}{\partial x^k} \left( \frac{\partial\mathcal{L}} {\partial u_{kx}(x,t)} \right),
\label{eq:ELlocal}
\end{align}
and
\begin{align}
\mathcal{E}_{-x}(u) = 
\frac{\partial\mathcal{L}}{\partial u(-x,t)} - \frac{\partial}{\partial t} \left( \frac{\partial\mathcal{L}} {\partial u_t(-x,t)} \right)
+
\sum_{k=1}^{n} \frac{\partial^k}{\partial x^k} \left( \frac{\partial\mathcal{L}} {\partial u_{kx}(-x,t)} \right).
\label{eq:ELreflected}
\end{align}
The second integral in Eq.~\eqref{eq:variationSeparated} is independently transformed into an integral over the variation $\delta u(x,t)$ by performing the reflection transformation $x\rightarrow -x$. Moreover, since $dx=-d(-x)$, the reversal of the integration limits restores the original orientation of the integral, giving
\begin{equation}
\iint_{-\infty}^\infty  dx\,dt\, \mathcal{E}_{-x}(u)\, \delta u(-x,t) =
\iint_{-\infty}^\infty  dx\,dt\, \left[ \mathcal{E}_{-x}(u) \right]_{x\rightarrow -x} \delta u(x,t).
\end{equation}
Therefore, the variation of the action can be written entirely in terms of the variation $\delta u(x,t)$,

\begin{equation}
\delta S = \iint_{-\infty}^\infty dx\,dt \, \Bigg[ \mathcal{E}_{x}(u) + \left[ \mathcal{E}_{-x}(u) \right]_{x\rightarrow -x} \Bigg] \delta u(x,t).
\label{eq:variationFinal}
\end{equation}
Since $\delta u(x,t)$ is arbitrary throughout the integration domain, the fundamental lemma of the calculus of variations implies
\begin{equation}
\mathcal{E}_{x}(u) + \left[ \mathcal{E}_{-x}(u) \right]_{x\rightarrow -x} =0.
\label{eq:GeneralizedEL1}
\end{equation}
This equation constitutes the generalized Euler-Lagrange equation for Lagrangian densities depending explicitly on both the field and its spatially reflected counterpart. In the local limit, where the Lagrangian density depends only on the field evaluated at $x$, the reflected contribution vanishes and Eq.~\eqref{eq:GeneralizedEL1} reduces to the conventional Euler-Lagrange equation. Equivalently, the generalized Euler–Lagrange equation associated with variations with respect to $u(-x,t)$
and can be written as 
\begin{equation}
\mathcal{E}_{-x}(u) + \left[ \mathcal{E}_{x}(u) \right]_{x\rightarrow -x} =0.
\label{eq:GeneralizedEL2}
\end{equation}
Motivated by the generalized Euler-Lagrange equation, we define the generalized functional derivative of an arbitrary functional
$F[u]$ by
\begin{align}
\frac{\delta F[u]}{\delta u(x,t)}= 
& \frac{\partial F}{\partial u(x,t)} - \frac{\partial}{\partial t} \left( \frac{\partial F} {\partial u_t(x,t)} \right) + \sum_{k=1}^{n} (-1)^k \frac{\partial^k}{\partial x^k} \left( \frac{\partial F} {\partial u_{kx}(x,t)} \right)
\nonumber\\
& + \Bigg[ \frac{\partial F}{\partial u(-x,t)} - \frac{\partial}{\partial t} \left( \frac{\partial F} {\partial u_t(-x,t)} \right) + \sum_{k=1}^{n} \frac{\partial^k}{\partial x^k} \left( \frac{\partial F} {\partial u_{kx}(-x,t)} \right) \Bigg]_{x\rightarrow -x}
\label{eq:GFD}
\end{align}
The generalized functional derivative introduced above allows the canonical Poisson bracket to be extended in a natural way to spatially nonlocal field theories. Therefore, for functionals $F[\phi,\pi]$ and $G[\phi,\pi]$, 
\begin{equation}
\{F,G\}
=
\sum_{\alpha=1}^{N}
\int dx
\left(
\frac{\delta F}{\delta \phi^\alpha(x,t)}
\frac{\delta G}{\delta \pi_\alpha(x,t)}
-
\frac{\delta F}{\delta \pi_\alpha(x,t)}
\frac{\delta G}{\delta \phi^\alpha(x,t)}
\right),
\label{eq:GeneralizedPoissonBracket}
\end{equation}
where $\phi^\alpha(x,t)$ and $\pi_\alpha(x,t)$ denote a set of canonical fields and their conjugate momenta, and $\alpha=1,2,3, \dots, n$ labels the canonical field components. 

The generalized Hamiltonian formalism preserves the algebraic structure of the standard Hamiltonian theory. The essential modification is the replacement of ordinary functional derivatives by the generalized functional derivative defined in Eq.~\eqref{eq:GFD}. Consequently, every canonical construction, including canonical momenta, Hamilton's equations, Poisson brackets, and the Dirac-Bergmann constraint algorithm, retains its standard algebraic form and reduces smoothly to its local counterpart in the appropriate limit.

The formalism developed above can now be applied to specific spatially nonlocal equations. We first consider the Ablowitz-Musslimani equation, which has an explicit parity symmetry. We then consider the LN1 and LN2 equations, which possess spatially nonlocal Lagrangian structures without parity symmetry. For each case, we derive the generalized Euler-Lagrange equations, identify the canonical variables and primary constraints, construct the canonical and total Hamiltonians, and verify that the generalized Hamiltonian formalism reproduces the corresponding equations of motion.
\section{Hamiltonian Formalism of the Ablowitz-Musslimani Equation}\label{AM formalism}

The Ablowitz-Musslimani nonlocal NLS equation is given by~\cite{ablowitz2013integrable}
\begin{equation}
i q_t-q_{xx}+2\sigma q^2q^{*}(-x)=0,
\label{eq:AM-NLS}
\end{equation}
where, for convenience, we use the notation
\begin{align}
    q(x,t) &\equiv q ~,
    \\
    q(-x,t)&\equiv q(-x)~.
\end{align}
We construct the following Lagrangian density for Eq.~\eqref{eq:AM-NLS}, which is explicitly written in terms of the complex fields $q$ and $q^*$.
\begin{align}\label{complex lagrangian}
    \mathcal{L}_{\rm AM}= &-\frac{i}{4}\left[ q^*_t(-x) q- q_t(-x) q^*  +q^*_tq(-x)-q_tq^*(-x) \right]  \nonumber \\
    &- \frac{1}{2}\left[q_x q_x^*(-x)+q_x(-x) q_x^* -\sigma \left( q^2(-x){q^*}^2 +q^2{q^*}^2(-x) \right) \right]
\end{align}
To the best of our knowledge, Eq.~\eqref{complex lagrangian} is the first standard Lagrangian density formulated directly in terms of the complex fields $q(x,t)$ and $q^{*}(-x,t)$. 

Although the Lagrangian density Eq.~\eqref{complex lagrangian} is written in terms of the complex fields $q$ and $q^{*}$ and contains explicit factors of the imaginary unit $i$, it is in fact real. This becomes evident after decomposing the complex field into its real and imaginary components by
\begin{equation}
\begin{aligned}
q(x) &= \phi(x)+i\psi(x),\\
q^{*}(-x) &= \phi(-x)-i\psi(-x),
\end{aligned}
\label{eq:decomposition}
\end{equation}
which transforms the Lagrangian density to
\begin{align}\label{Lagrangian} \nonumber
    \mathcal{L}_{\rm AM} =& \frac{1}{2}\left[ \phi_t 
 \psi(-x)+\phi_t(-x)\psi-\phi\psi_t(-x)-\phi(-x)\psi_t   \right] -\phi_x\phi_x(-x)-\psi_x\psi_x(-x)
 \\&
 +\sigma\left[\left(  \phi^2 -\psi^2 \right)\left(  \phi^2(-x) -\psi^2(-x) \right)+ 4 \phi \psi \phi(-x)\psi(-x)  \right].
\end{align}
The resulting Lagrangian density is manifestly real and is expressed entirely in terms of the real fields $\phi$ and $\psi$. Likewise, the decomposition Eq.~\eqref{eq:decomposition} transforms Eq.~\eqref{eq:AM-NLS} into the coupled system

\begin{align}\label{eom-psi}
    &\psi_t +\phi_{xx}-2\sigma \left[ \phi^2\phi(-x) +2\phi \psi \psi(-x)-\psi^2\phi(-x) \right]=0,
    \\ \label{eom-phi}
    &\phi_t -\psi_{xx}+2\sigma \left[ -\phi^2\psi(-x) +2\phi \psi \phi(-x)+\psi^2\psi(-x) \right]=0.
\end{align}

As expected, the generalized Euler-Lagrange equations \eqref{eq:GeneralizedEL1} and \eqref{eq:GeneralizedEL2} associated with the real Lagrangian density Eq.~\eqref{Lagrangian} reproduce the coupled equations of motion. Specifically, variations with respect to $\phi(-x)$ and $\psi(-x)$ yield Eqs.~\eqref{eom-psi} and \eqref{eom-phi}, respectively, while variations with respect to $\phi(x)$ and $\psi(x)$ yield their reflected counterparts under the reflection $x\rightarrow -x$.

Having established the Lagrangian formulation and verified that it reproduces the nonlocal NLS equation through the generalized Euler-Lagrange equations, we now construct its Hamiltonian. The Lagrangian density Eq.~\eqref{Lagrangian} is linear in the generalized velocities, implying that the Hessian matrix 
\begin{equation}\label{Hessian matrix}
    M_{ij}= \frac{\delta^2 \mathcal{L}_{\rm AM}}{\delta \dot{v}_i \delta \dot{v}_j},
\end{equation}
where $v=(\phi,\psi)$, and its determinant vanishes identically. Consequently, the system must be treated within the Dirac-Bergmann algorithm (DBA) of constrained Hamiltonian systems~\cite{Dirac1950, YNutku_1983,Nutku1984,Rothe2010, Deriglazov2010,salisbury2017, Lusanna2018,filiz2018,filiz2020,gumral2022, pazarci2026, russkov2026}. The total Hamiltonian $H$ is constructed by adding the contribution of the constraints, $\mathcal{H}_c$, to the canonical Hamiltonian density, $\mathcal{H}_L$.
\begin{equation}
    H_{\rm AM} = \int \mathcal{H}_{\rm AM}\:dx =\int (\mathcal{H}_L+\mathcal{H}_c)\:dx\,. 
\end{equation}
Since the Hessian matrix vanishes, the generalized velocities cannot be solved uniquely in terms of the canonical momenta. Consequently, the definitions of the canonical momenta will lead to two independent primary constraints. 

The generalized canonical momenta associated with the fields $\phi$ and $\psi$ are
\begin{align}
\pi_\phi &= \frac{\delta\mathcal{L}_{\rm AM}} {\delta\phi_t} = \psi(-x), 
\label{momenta phi} \\
\pi_\psi &= \frac{\delta\mathcal{L}_{\rm AM}} {\delta\psi_t} = -\phi(-x),
\label{momenta psi}
\end{align}
which give rise to the two independent primary constraints
\begin{align}
c_1 &= \pi_\phi-\psi(-x), \\
c_2 &= \pi_\psi+\phi(-x).
\end{align}
The constraints' contribution to the total Hamiltonian is therefore 
\begin{equation}
\mathcal H_{c} = \lambda_1(x) \Big(\pi_\phi-\psi(-x)\Big) +\lambda_2(x) \Big( \pi_\psi+\phi(-x) \Big).
\end{equation}
Since the system enjoys a parity symmetry, one can use the spatially reflected version of canonical momenta, i.e., $\pi_\phi(-x)$ and $\pi_\psi(-x)$ to construct the constraints' Hamiltonian density. The canonical Hamiltonian density $H_L$ can be constructed through the Legendre transformation,
\begin{equation}
\mathcal H_{L} =  \pi_{\phi}\phi_t   + \pi_{\psi}\psi_t  - \mathcal L.
\end{equation}
It is convenient to use the spatially symmetric nature of the system and express the canonical Hamiltonian density $H_L$, in terms of the local fields and their reflected counterparts in this construction 
\begin{equation}
\mathcal H_{L} = \frac12 \left[ \pi_{\phi}\phi_t + \pi_{\phi}(-x)\phi_t(-x) 
+ \pi_{\psi}\psi_t + \pi_{\psi}(-x)\psi_t(-x) \right] - \mathcal L,
\end{equation}
Finally, the sum of $H_c$ and $H_L$ gives the total Hamiltonian density as
\begin{align}\nonumber
    \mathcal{H}_{\rm AM} =&~ \phi_x\phi_x(-x)+\psi_x\psi_x(-x)
 -\sigma\left[ \left(  \phi^2 -\psi^2 \right)\big(  \phi^2(-x) -\psi^2(-x) \big)+ 4 \phi \psi \phi(-x)\psi(-x)  \right]
\nonumber \\
&  +
 \lambda_1(x) \Big(\pi_\phi-\psi(-x)\Big) +\lambda_2(x) \Big( \pi_\psi+\phi(-x) \Big)
\end{align}
where $\lambda_1(x)$ and $\lambda_2(x)$ are Lagrange multipliers to be determined. The Lagrangian naturally leads to the canonical Poisson bracket relations
\begin{align}
\{\phi(x),\pi_\phi(y)\} = -\{\pi_\phi(x),\phi(y)\} =\delta(x-y) ,\\
\{\psi(x),\pi_\psi(y)\} = -\{\pi_\psi(x),\psi(y)\} = \delta(x-y).
\end{align}
The preservation of the primary constraints under time evolution requires
\begin{equation}\label{consistency conditions}
\dot c_i = \{c_i,H_{\rm AM}\} \approx0, \qquad i=1,2~,
\end{equation}
which determine the Lagrange multipliers as
\begin{align}
    \lambda_1(x) &= \psi_{xx}-2\sigma\left[ -\psi(-x) \left( \phi^2-\psi^2 \right)+2\phi(-x) \phi\psi
    \right],
    \\
    \lambda_2(x) &= -\phi_{xx}+2\sigma\left[ \phi(-x) \left( \phi^2-\psi^2 \right)+2\psi(-x) \phi\psi
    \right],
\end{align}
where the reflected expressions have been mapped back to the coordinate $x$ through the transformation $x\rightarrow -x$. Note that introducing the multipliers and the constraints in the reflected space, such as $\lambda_1(-x)$, $\lambda_2(-x)$, $c_1(-x)$, and $c_2(-x)$, does not affect the results, since they fix themselves accordingly during the consistency conditions Eq.~\eqref{consistency conditions}. Substituting the expressions for the constraints into the total Hamiltonian gives
\begin{align}\label{Hamiltonian with momenta}
    \mathcal{H}_{\rm AM}=&~ \sigma\left(\phi^2 - \psi^2 \right)\left[\phi^2(-x) -\psi^2(-x) \right] 
    + 4 \sigma \phi \psi \phi(-x)\psi(-x) \nonumber \\&
    + \pi_\phi \left(\psi_{xx}-2\sigma \left[-\psi(-x) \left( \phi^2-\psi^2 \right)+2\phi \psi \phi(-x) \right] \right) 
    \nonumber \\&  
    + \pi_\psi\left( -\phi_{xx}+2\sigma\left[ \phi(-x) \left( \phi^2-\psi^2 \right)+2\phi \psi \psi(-x) \right]\right) .
\end{align}

The Hamilton equations of motion are then obtained from
\begin{align}
    (v_i)_t &=\{v_i,H_{\rm AM}\},
    \\
    (\pi_{v_i})_t&=\{\pi_{v_i},H_{\rm AM}\},
\end{align}
where $\pi_{v_i}=(\pi_{\phi}, \pi_{\psi})$, which give rise to the following equations. 
\begin{align}\label{Hamilton eom for phi}
    \phi_t=~ & \psi_{xx} - 2\sigma \left[ -\phi^2\psi(-x) +2\phi \psi \phi(-x)+\psi^2\psi(-x) \right],
    \\\label{Hamilton eom for psi}
    \psi_t =~ &-\phi_{xx} + 2\sigma \left[ \phi^2\phi(-x) +2\phi \psi \psi(-x)-\psi^2\phi(-x) \right],
    \\ \label{Hamilton eom for pi_phi}
    (\pi_\phi)_t =~ & 4\sigma\left[-\phi\left(\phi^2(-x)-\psi^2(-x) \right)-2\psi\phi(-x)\psi(-x) \right]-4\sigma\pi_\phi (\phi \psi(-x) -\psi\phi(-x))+4\sigma\pi_\phi(-x)\phi(-x)\psi(-x) 
    \\& \nonumber
    +(\pi_\psi)_{xx}-2 \sigma\pi_\psi(-x)\left(\phi^2(-x)-\psi^2(-x) \right) -4\sigma\pi_\psi(\psi\psi(-x)+\phi\phi(-x)),
    \\ \label{Hamilton eom for pi_psi}
    (\pi_\psi)_t=~ & 4 \sigma\left[ \psi\left(\phi^2(-x)-\psi^2(-x) \right)-2\phi \phi(-x)\psi(-x) \right] -(\pi_\phi)_{xx}-2\sigma\pi_\phi(-x)\left(\phi^2(-x)-\psi^2(-x) \right)
    \\& \nonumber
    +4\sigma \pi_\phi(\psi\psi(-x)+\phi\phi(-x)) -4\sigma \pi_\psi(\phi \psi(-x)-\psi \phi(-x))-4\sigma\pi_\psi(-x)\phi(-x)\psi(-x).
\end{align}
Substituting the canonical momenta, Eqs.~\eqref{momenta phi} and \eqref{momenta psi}, into the resulting equations for the canonical momenta reproduces reflected counterparts of Eqs.~\eqref{eom-phi} and \eqref{eom-psi}. Therefore, the generalized Hamiltonian formalism reproduces exactly the generalized Euler-Lagrange equations derived previously. 

Eliminating the canonical momenta through the primary constraints and integrating by parts, the Hamiltonian density Eq.~\eqref{Hamiltonian with momenta} assumes the compact form
\begin{align}\label{substituted hamiltonian}
    \mathcal{H}_{\rm AM}= \phi_x \phi_x(-x)+\psi_x \psi_x(-x)-\sigma\left[\left(  \phi^2 -\psi^2 \right)\left(  \phi^2(-x) -\psi^2(-x) \right)+ 4 \phi \psi \phi(-x)\psi(-x)  \right].
\end{align}
Recombining the real fields into the original complex field gives the Hamiltonian 
\footnote{The authors in \cite{Ablowitz2018} presented the Hamiltonian for this system as
 \begin{align}
     H&= \int_{-\infty}^{\infty} dx\left[- q_x q_x^*(-x) -\sigma q^2{q^*}^2(-x)\right]~,\quad \sigma=\pm 1~.
 \end{align}
However, it is important to note that they do not use the generalized functional derivative \eqref{eq:GFD} that is introduced in this paper while computing \eqref{Complex Hamilton eom}.}
\begin{equation}\label{AME hamiltonian written in complex fields}
    H_{\rm AM}=\int_{-\infty}^{\infty} dx\left[\frac{1}{2}\Big( q_x q_x^*(-x)+q_x(-x) q_x^* \Big) -\frac{\sigma}{2} \Big( q^2(-x){q^*}^2 +q^2{q^*}^2(-x)\Big)\right],
\end{equation}
which generates the nonlocal NLS equation through
\begin{equation}\label{Complex Hamilton eom}
    i q_t = \frac{\delta H_{\rm AM}}{\delta q^*(-x)}.
\end{equation}
Among the infinitely many integrals of motion of this system, Ablowitz and Musslimani give the following three~\cite{Ablowitz2018}:
\begin{align}
    \mathcal{I}_1 &=\int_{-\infty}^{\infty} dx \, qq^*(-x)~,
    \\
    \mathcal{I}_2 &=  \int_{-\infty}^{\infty} dx\left[ q_x q^*(-x) + qq_x^*(-x)\right],
    \\
    \mathcal{I}_3 &=\int_{-\infty}^{\infty} dx\left[ q_x q_x^*(-x) -\sigma q^2{q^*}^2(-x)\right].
\end{align}
To verify that these integrals of motion are in involution with the Hamiltonian, it is convenient to express them in terms of the real fields defined in Eq.~\eqref{eq:decomposition}. We therefore write
\begin{align}\label{IEone}
    \mathcal{I}_1 = \int_{-\infty}^{\infty} dx & \,  \left[ \phi \phi(-x)+\psi \psi(-x) \right] + i\left[ \psi \phi(-x) -\phi \psi(-x) \right],
    \\ \label{IEtwo}
    \mathcal{I}_2 = \int_{-\infty}^{\infty}  dx & \,\Big\{ \left[ \phi_x \phi(-x)+\phi \phi_x(-x) + \psi_x \psi(-x)+\psi \psi_x(-x) \right] \nonumber \\&
    + i \left[ \psi_x\phi(-x)+\psi\phi_x(-x)-\phi_x\psi(-x)-\phi\psi_x(-x) \right]\Big\},
    \\\label{IEthree}
    \mathcal{I}_3 = \int_{-\infty}^{\infty} dx & \, \Big\{\Big[ \phi_x\phi_x(-x) + \psi_x\psi_x(-x)  -\sigma \big[ \big(\phi^2-\psi^2\big)\big(\phi^2(-x)-\psi^2(-x)\big)+4\phi\psi\phi(-x)\psi(-x) \big] \Big] \nonumber \\& 
    + i \Big[ \psi_x\phi_x(-x)-\phi_x\psi_x(-x) -2 \sigma \big[\phi\psi\big( \phi^2(-x)-\psi^2(-x) \big) -  \phi(-x)\psi(-x) \big(  \phi^2-\psi^2 \big)\big] \Big]\Big\}.
\end{align}
Using the generalized Poisson bracket defined in Eq.~\eqref{eq:GeneralizedPoissonBracket}, it can be shown that 
\begin{equation}
\{\mathcal I_i,H_{\rm AM}\}\approx0,
\qquad i=1,2,3.
\end{equation}
Note that the imaginary parts of $\mathcal{I}_i$'s in Eqs.\eqref{IEone}-\eqref{IEthree}, are antisymmetric under the parity transformation and therefore vanish upon integration. Equivalently, their generalized functional derivatives with respect to the fields vanish. Thus, only the real parts of $\mathcal{I}_i$'s contribute to the Poisson brackets.

The Hamiltonian, Eq.~\eqref{AME hamiltonian written in complex fields}, can also be
expressed in terms of the third integral of motion as
\begin{equation}
H_{\rm AM}
=\frac{1}{2}\left(
\mathcal{I}_3+
\left.\mathcal{I}_3\right|_{x\rightarrow -x}
\right).
\end{equation}
Consequently, the generalized Hamiltonian formalism not only reproduces the correct equations of motion but also recovers the Hamiltonian associated with the integrable structure of the nonlocal NLS equation.

\section{Hamiltonian formalism of the LN1 Equation}\label{LN1 formalism}
We now apply the generalized Hamiltonian formalism developed in Section~\ref{NL formalism} to the first Lagrangian nonlocal NLS equation (LN1) introduced by Velasco-Juan and Fujioka~\cite{velasco2022}. The equation is given by
\begin{equation}\label{LN1 equation}
i u_t -u_{xx} -\frac{1}{2}u^2u^{*}(-x) -\left[ |u|^2 +\frac{1}{2}|u(-x)|^2 \right]u(-x) =0.
\end{equation}
Here $u\equiv u(x,t)$ and $u(-x) \equiv u(-x,t)$. In the original complex representation, the Lagrangian density is given by
\begin{equation}
\mathcal{L}_{\mathrm{LN1}} = -\operatorname{Im}\!\left(u^{*}u_{t}\right) 
+ |u_{x}|^{2} - |u|^{2} \operatorname{Re} \!\left[ u(x)u^{*}(-x) \right].
\label{LN1_Lagrangian}
\end{equation}
Using the field decomposition
\begin{equation}
\begin{aligned}\label{eq:decomposition-ln1}
u(x)&=\phi(x)+i\psi(x),\\
u^{*}(-x)&=\phi(-x)-i\psi(-x),
\end{aligned}
\end{equation}
the Lagrangian density assumes the equivalent form
\begin{align}
\mathcal{L}_{\mathrm{LN1}} ={}& -\phi\psi_t +\psi\phi_t +\phi_x^2 +\psi_x^2 
-(\phi^2+\psi^2) \left[ \phi\,\phi(-x) + \psi\,\psi(-x) \right],
\end{align}
in terms of real field components. 

Applying the generalized Euler-Lagrange equations \eqref{eq:GeneralizedEL1} and \eqref{eq:GeneralizedEL2} to this Lagrangian density produces the coupled equations of motion
\begin{align}
2\phi_t - 2\psi_{xx} - \phi^2\psi(-x) - 2\phi\psi\phi(-x) - 3\psi^2\psi(-x) - \phi^2(-x)\psi(-x) - \psi^3(-x) = 0, 
\label{EL1-LN1}\\
2\psi_t + 2\phi_{xx} + 3\phi^2\phi(-x) + 2\phi\psi\,\psi(-x) + \psi^2\phi(-x) + \phi^3(-x) + \phi(-x)\psi^2(-x) = 0,
\label{EL2-LN1} 
\end{align}
where variations with respect to $\phi(x)$ and $\psi(x)$ yield the above equations, while variations with respect to $\phi(-x)$ and $\psi(-x)$ produce their reflected counterparts. Euler-Lagrange equations \eqref{EL1-LN1} and \eqref{EL2-LN1} can also be obtained by decomposing Eq.~\eqref{LN1 equation} into its real and imaginary parts.

Since the Lagrangian density is linear in the generalized velocities, the Hessian matrix vanishes identically. Consequently, the system is singular and the generalized Dirac-Bergmann algorithm applies. The generalized canonical momenta
\begin{align}
 \pi_{\phi} &= \frac{\partial\mathcal{L}_{\rm LN1}}{\partial\phi_t} = \psi,
\label{pi_phi-ln1} \\
 \pi_{\psi} &= \frac{\partial\mathcal{L}_{\rm LN1}}{\partial\psi_t} = -\phi.
\label{pi_psi-ln1}
\end{align}
immediately give rise to the primary constraints
\begin{align}
c_{1} = \pi_{\phi} - \psi,
\label{chiln1}\\
c_{2} = \pi_{\psi} + \phi.
\label{chiln2}
\end{align}
The total Hamiltonian is therefore constructed as
\begin{equation}
H_{\rm LN1}=\int\mathcal H_{\rm LN1}\,dx,
\end{equation}
where
\begin{align}
\mathcal H_{\rm LN1} = \mathcal H_L + \mathcal H_c =~ 
&  -\phi_x^2 -\psi_x^2 +\phi^3\phi(-x) +\phi^2\psi\,\psi(-x)  +\phi\psi^2\phi(-x) +\psi^3\psi(-x) \nonumber \\
&  +\lambda_{1}(\pi_{\phi}-\psi) +\lambda_{2}(\pi_{\psi}+\phi).
\label{HT_1}
\end{align}
The consistency conditions of the primary constraints,
\begin{equation}
\dot c_i = \{c_i,H_{\rm LN1}\} \approx0, \qquad i=1,2~,
\end{equation}
determines the Lagrange multipliers as
\begin{align}
\lambda_{1} &= \psi_{xx} +\frac12\phi^{2}\psi(-x) +\phi\psi\,\phi(-x) +\frac32\psi^{2}\psi(-x) +\frac12\phi^{2}(-x)\psi(-x) +\frac12\psi^{3}(-x) ,\\
\lambda_{2} &= -\phi_{xx} -\frac{3}{2}\phi^{2}\phi(-x) -\phi\psi\,\psi(-x) -\frac{1}{2}\psi^{2}\phi(-x) -\frac{1}{2}\phi^{3}(-x) -\frac{1}{2}\phi(-x)\psi^{2}(-x).
\end{align}
Substituting these expressions into the total Hamiltonian yields
\begin{align}
\mathcal H_{\rm LN1} ={}& -\frac12\phi^3\phi(-x) -\frac12\phi^2\psi\,\psi(-x) -\frac12\phi\psi^2\phi(-x) -\frac12\psi^3\psi(-x) \nonumber \\
& -\frac12\phi\,\phi^3(-x) -\frac12\phi^2(-x)\psi\,\psi(-x) -\frac12\phi\,\phi(-x)\psi^2(-x) -\frac12\psi\,\psi^3(-x) \nonumber \\
& +\pi_\phi \Bigg[ \psi_{xx} +\frac12\phi^{2}\psi(-x) +\phi\psi\,\phi(-x) +\frac32\psi^{2}\psi(-x)  +\frac12\phi^{2}(-x)\psi(-x) +\frac12\psi^{3}(-x) \Bigg] \nonumber  \\
& -\pi_\psi \Bigg[ \phi_{xx} +\frac32\phi^{2}\phi(-x) +\phi\psi\,\psi(-x) +\frac12\psi^{2}\phi(-x) +\frac12\phi^{3}(-x) +\frac12\phi(-x)\psi^{2}(-x) \Bigg].
\end{align}
The corresponding Hamilton equations reproduce exactly the generalized Euler-Lagrange equations \eqref{EL1-LN1} and \eqref{EL2-LN1}. Therefore, the generalized Hamiltonian formalism developed in Section~\ref{NL formalism} provides a complete Hamiltonian description of the LN1 equation too.

Furthermore, using the primary constraints Eqs.~\eqref{pi_phi-ln1} and \eqref{pi_psi-ln1}, the total Hamiltonian density simplifies to
\begin{equation}
\mathcal H_{\rm LN1} = -\phi_x^2 - \psi_x^2 + (\phi^2+\psi^2) \left[ \phi\phi(-x) +\psi\psi(-x) \right].
\label{HT_reduced_factorized}
\end{equation}
Recombining the real fields according to Eq.~\eqref{eq:decomposition-ln1}, the total Hamiltonian can be expressed in terms of the original complex field as
\begin{equation}
H_{\rm LN1} = \int_{-\infty}^{\infty} dx \left\{-|u_{x}|^2
+ \frac{1}{2}|u|^{2}\left[uu^{*}(-x)+u^*u(-x)\right]\right\}.
\label{ln1_complex}
\end{equation}

\section{Hamiltonian Formalism of the LN2 Equation}\label{LN2 formalism}

We finally apply the generalized Hamiltonian formalism developed in Section~\ref{NL formalism} to the second Lagrangian nonlocal NLS equation (LN2) introduced by Velasco-Juan and Fujioka~\cite{velasco2022}. As in the LN1 case, the equation possesses a Lagrangian formulation, while its Hamiltonian structure is obtained here systematically within the generalized Hamiltonian formalism by using the Dirac-Bergmann algorithm.

The LN2 equation is given by
\begin{equation}\label{LN2 equation}
i u_t-u_{xx}-\left[\,|u|^2+|u(-x)|^2\,\right]u=0.
\end{equation}
The corresponding complex Lagrangian density proposed in \cite{velasco2022} is
\begin{equation}
\mathcal{L}_{\mathrm{LN2}} = -\operatorname{Im}(u^{*}u_t) + |u_x|^2 - \frac12|u|^4 - \frac12|u|^2|u(-x)|^2.
\end{equation}
Introducing the decomposition Eq.~\eqref{eq:decomposition-ln1}, the Lagrangian density assumes the equivalent real form
\begin{equation}
\mathcal{L}_{\mathrm{LN2}} = -\phi\psi_t +\psi\phi_t +\phi_x^2 +\psi_x^2 -\frac12(\phi^2+\psi^2)^2 
-\frac12(\phi^2+\psi^2) \left[ \phi(-x)^2+\psi(-x)^2 \right].
\end{equation}
The generalized Euler-Lagrange equations  \eqref{eq:GeneralizedEL1}  of $\mathcal{L}_{\mathrm{LN2}} $ result in the following coupled equations of motion
\begin{align}
\psi_t + \phi_{xx} + \phi(\phi^2+\psi^2) + \phi \left[ \phi^2(-x)+\psi^2(-x) \right] =0, \label{EL1-LN2}\\
\phi_t - \psi_{xx} - \psi(\phi^2+\psi^2) - \psi \left[ \phi^2(-x)+\psi^2(-x) \right] =0,\label{EL2-LN2}
\end{align}
whereas the generalized Euler-Lagrange equations \eqref{eq:GeneralizedEL2} of $\mathcal{L}_{\mathrm{LN2}} $ produce their spatially reflected counterparts. As it was the case for AM and LN1 equations, the decomposition of the Eq.~\eqref{LN2 equation} also produces the same equations \eqref{EL1-LN2} and \eqref{EL2-LN2}.

The Hessian matrix of $\mathcal{L}_{\mathrm{LN2}} $ vanishes since the Lagrangian density is linear in time derivatives. Therefore, the generalized Dirac-Bergmann algorithm should be considered to obtain a consistent Hamiltonian formulation. The generalized canonical momenta are
\begin{align} 
\pi_\phi &= \psi,\label{pi_phi-ln2}\\
\pi_\psi &= -\phi\label{pi_psi-ln2},
\end{align}
which immediately give rise to the primary constraints
\begin{align}
c_1 &= \pi_{\phi}-\psi,\\
c_2 &= \pi_{\psi}+\phi.
\end{align}
The total Hamiltonian is constructed as
\begin{equation}
H_{\rm LN2}=\int\mathcal{H}_{\rm LN2}\,dx,
\end{equation}
where
\begin{align}
\mathcal{H}_{\rm LN2} = \mathcal{H}_L + \mathcal{H}_c =
& -\phi_x^2 -\psi_x^2 +\frac12(\phi^2+\psi^2)^2 +\frac12(\phi^2+\psi^2) \left[ \phi^2(-x)+\psi^2(-x) \right] \nonumber \\
& + \lambda_1 \left( \pi_\phi-\psi \right) + \lambda_2 \left( \pi_\psi+\phi \right).
\end{align}
The preservation of the primary constraints under time evolution,
\begin{equation}
\dot{c}_{i}
=
\{c_i,H_{\rm LN2}\}
\approx0,
\qquad
i=1,2~,
\end{equation}
determines the Lagrange multipliers,
\begin{align}
\lambda_1 &= \psi_{xx} + \psi(\phi^2+\psi^2) + \psi \left[ \phi^2(-x)+\psi^2(-x) \right], \\
\lambda_2 &= -\phi_{xx} - \phi(\phi^2+\psi^2) - \phi \left[ \phi^2(-x)+\psi^2(-x) \right].
\end{align}
Substituting these expressions into the total Hamiltonian yields the complete Hamiltonian density
\begin{align}
\mathcal H_{\rm LN2} =~
& -\frac12(\phi^2+\psi^2)^2 -\frac12(\phi^2+\psi^2) \left[ \phi^2(-x)+\psi^2(-x) \right] \nonumber \\
& +\pi_\phi \Bigg[ \psi_{xx} +\psi(\phi^2+\psi^2) +\psi \left( \phi^2(-x)+\psi^2(-x) \right) \Bigg] \nonumber \\ \label{Final Total Hamiltonian of LN2}
& -\pi_\psi \Bigg[ \phi_{xx} +\phi(\phi^2+\psi^2) +\phi \left( \phi^2(-x)+\psi^2(-x) \right) \Bigg].
\end{align}
The Hamilton equations generated by this Hamiltonian reproduce exactly the generalized Euler-Lagrange equations \eqref{EL1-LN2} and \eqref{EL2-LN2}, while the equations for the canonical momenta reduce to their corresponding spatially reflected counterparts.

Using the primary constraints to substitute momenta, Eqs.~\eqref{pi_phi-ln2} and \eqref{pi_psi-ln2}, into Eq.~\eqref{Final Total Hamiltonian of LN2}, the total Hamiltonian density simplifies to
\begin{align}
\mathcal H_{\rm LN2} = -\phi_{x}^2-\psi_{x}^2 + \frac12(\phi^2+\psi^2)^2 +\frac12(\phi^2+\psi^2) \left[ \phi^2(-x)+\psi^2(-x) \right].
\end{align}
Finally, recombining the real fields according to Eq.~\eqref{eq:decomposition-ln1}, the Hamiltonian density can be expressed in terms of the original complex field as
\begin{equation}
H_{\rm LN2} = \int_{-\infty}^{\infty} dx \left[-|u_x|^2
+\frac{1}{2}|u|^4
+\frac{1}{2}|u|^2|u(-x)|^2\right].
\label{ln2_reduced_complex}
\end{equation}
\section{Summary and Discussion}\label{Discussion}
In this work, we developed a generalized Hamiltonian formalism for spatially nonlocal field theories and applied it to three representative nonlocal NLS equations. Starting from a generalized variational principle, we derived the corresponding generalized Euler-Lagrange equations, introduced a generalized functional derivative, and showed that the resulting formalism provides a consistent Hamiltonian description of the AM, LN1, and LN2 equations using the Dirac-Bergmann algorithm.

We also presented, to the best of our knowledge, the first standard Lagrangian density written for the AM system. Although this Lagrangian density is constructed directly in terms of the complex fields and contains explicit factors of the imaginary unit $i$, it is in fact real. This becomes evident after decomposing the complex field into its real and imaginary components. The AM system also possesses an explicit parity symmetry, which is manifested in the structure of its Lagrangian and Hamiltonian formulations.

Having established a Lagrangian for the AM equation, we constructed its complete Hamiltonian formulation and showed that the resulting Hamilton equations reproduce the equations obtained from the generalized Euler-Lagrange equations. The same formalism was then applied without modification to the LN1 and LN2 equations, demonstrating that it is not restricted to a single model but provides a unified framework for a broader class of spatially nonlocal systems. This framework can also be extended to other types of space and time nonlocalities such as $(x_0-x,\pm t)$ \cite{ablowitz2021}, $(x \pm x_0,\pm t)$ \cite{fujioka2019}, $(\pm x,-t)$ \cite{Ablowitz2018, chen2018}, etc. We expect that these extensions will provide a useful foundation for further studies of symmetries, conservation laws, canonical quantization, and the geometric structure of nonlocal field theories.

The nonlinear Schrödinger equation arises in a wide variety of contexts in modern theoretical and mathematical physics (see, e.g. \cite{Nekrasov:2009ui,Nian2017, kaluc2022,koch2022generalized}). It is therefore natural to ask whether meaningful nonlocal generalizations can be constructed in these different settings, and what new physical and mathematical structures such extensions may reveal.

\section*{Acknowledgements}
\noindent 
The work of Ilmar Gahramanov was partially supported by the Boğaziçi University Research Fund under grant Y26.


\bibliography{NLE}

\end{document}